\documentclass[aip,amsmath,amssymb,app,reprint,superscriptaddress]{revtex4-2}

\usepackage{graphicx}
\usepackage{dcolumn}
\usepackage{bm}

\draft 

\begin{document}


\title{Wide-spectrum optical synthetic aperture imaging via spatial intensity interferometry} 



\author{Chunyan Chu}
\thanks{These authors contributed equally to this work.}
\affiliation{Beijing Key Laboratory for Precision Optoelectronic Measurement Instrument and Technology, Beijing, 100081, China}
\affiliation{School of Optics and Photonics, Beijing Institute of Technology, Beijing, 100081, China}

\author{Zhentao Liu}
\thanks{These authors contributed equally to this work.}
\affiliation{Shanghai Institute of Optics and Fine Mechanics, Chinese Academy of Sciences, 201800, China}
\affiliation{University of Chinese Academy of Sciences, Beijing, 100049, China}

\author{Mingliang Chen}
\email[]{Corresponding author: cml2008@siom.ac.cn.}
\affiliation{Shanghai Institute of Optics and Fine Mechanics, Chinese Academy of Sciences, 201800, China}
\affiliation{University of Chinese Academy of Sciences, Beijing, 100049, China}

\author{Xuehui Shao}
\affiliation{National Laboratory of Aerospace Intelligent Control Technology, Beijing, 100089, China}

\author{Yuejin Zhao}
\affiliation{Beijing Key Laboratory for Precision Optoelectronic Measurement Instrument and Technology, Beijing, 100081, China}
\affiliation{School of Optics and Photonics, Beijing Institute of Technology, Beijing, 100081, China}

\author{Shensheng Han}
\affiliation{Shanghai Institute of Optics and Fine Mechanics, Chinese Academy of Sciences, 201800, China}
\affiliation{Hangzhou Institute for Advanced Study, University of Chinese Academy of Sciences, Hangzhou, 310024, China}

\date{\today}

\begin{abstract}
High resolution imaging is achieved using increasingly larger apertures and successively shorter wavelengths. Optical aperture synthesis is an important high-resolution imaging technology used in astronomy. Conventional long baseline amplitude interferometry is susceptible to uncontrollable phase fluctuations, and the technical difficulty increases rapidly as the wavelength decreases. The intensity interferometry inspired by HBT experiment is essentially insensitive to phase fluctuations, but suffers from a narrow spectral bandwidth which results in a lack of detection sensitivity. In this study, we propose optical synthetic aperture imaging based on spatial intensity interferometry. This not only realizes diffraction-limited optical aperture synthesis in a single shot, but also enables imaging with a wide spectral bandwidth. And this method is insensitive to the optical path difference between the sub-apertures. Simulations and experiments present optical aperture synthesis diffraction-limited imaging through spatial intensity interferometry in a 100 $nm$ spectral width of visible light, whose maximum optical path difference between the sub-apertures reach $69.36\lambda$. This technique is expected to provide a solution for optical aperture synthesis over kilometer-long baselines at optical wavelengths.
\end{abstract}

\maketitle 

\section*{Introduction}
Intensity interferometry \cite{Brown1956} performed by Robert Hanbury Brown and Richard Twiss in 1956 not only ushered in the era of quantum optics, but also provided a new approach for long baseline imaging \cite{Twiss1969}. Because conventional long baseline amplitude interferometry is susceptible to phase fluctuations, its phase error must be corrected, and the technical difficulty increases rapidly as the wavelength decreases. In contrast to amplitude interferometry, intensity interferometry is essentially insensitive to phase fluctuations, therefore, it enables very long baselines to be observed at shorter optical wavelengths with higher resolution. From 1964 to 1972, the Narrabri Stellar Intensity Interferometer (NSII) first measured the angular diameters of 32 stars using two 6.5 $m$ reflectors on a circular track with a diameter of 188 $m$ \cite{Brown1974}. In HBT intensity interferometry, the intensity interferometer observes the second-order coherence of light by measuring the temporal correlations of the light intensity with different arrival times between photons recorded by different telescopes. Because the temporal coherence of light is determined by its bandwidth, the imaging bandwidth of this method is severely limited when recording the temporal intensity fluctuations of light. Owing to the lack of sensitivity \cite{Brown1974}, stellar intensity interferometry was almost abandoned in the 1970s \cite{Rivet2018}.
With the significant development of light detectors (such as single photon avalanche diodes, SPADs) and signal processing, stellar intensity interferometry has attracted the interest of researchers' again \cite{Nunez2011, Pilyavsky2017, Gori2021}. However, this lack of sensitivity has not been fundamentally solved.

Recent exciting advances in computational imaging have provided a new approach to far-field high-resolution imaging \cite{Gong2015, Bulbul2018, Liu2019, Bulbul2021, Wang2022}. A novel research reported by Bulbul et al. \cite{Bulbul2018} realized synthetic aperture-based imaging using coded phase reflectors distributed only along the boundaries of synthetic apertures. The image is obtained using cross-correlation between the detected system response to the object and system impulse response. However, this approach, which requiring a system impulse response, is limited to scenarios in which a guide-star is unobtainable. Inspired by ghost imaging using thermal light based on spatial intensity multi-correlation \cite{Liu2016}, Liu et al. \cite{Liu2019} proposed a lensless Wiener-Khinchin telescope based on spatial intensity autocorrelation, which can acquire an image via a spatial random phase modulator in a snapshot. However, the requirement for a large aperture monolithic spatial random phase modulator with a statistical distribution restricts this approach to up to aperture imaging ten or even hundred meters.

\begin{figure}[htbp]
\centerline{\includegraphics[width=1.0\columnwidth]{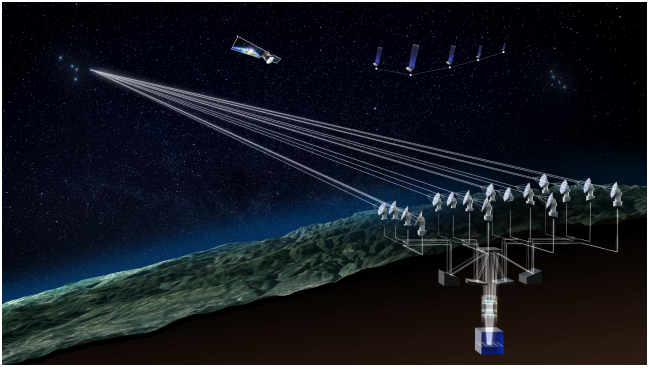}}
\caption{Application conception of optical synthetic aperture imaging via spatial intensity interferometry.}
 \label{Fig0}
\end{figure}

In this study, we present a technique that enables wide-spectrum optical synthetic aperture imaging via spatial intensity interferometry in a single shot by combining light from Wiener-Khinchin telescopes separated by baselines. We verified that it can achieve a coherent synthetic aperture  diffraction-limited resolution determined by the baseline length. Specifically, the energy spectral density of an object's image can be estimated using the intensity autocorrelation of the detected light, and the object's image can reconstructed using phase retrieval algorithms \cite{Fienup1982, Liu2015a, Shechtman2015, Shen2021}. As a proof of concept, we simulated and experimentally demonstrated wide-spectrum optical synthetic aperture imaging using a sub-aperture spatial random phase modulators(SRPMs) array. FIG. \ref {Fig0} depicts an application scenario for astronomical observation using optical synthetic aperture imaging via spatial intensity interferometry.

\section*{Principle}
\begin{figure}[htbp]
\centerline{\includegraphics[width=1.0\columnwidth]{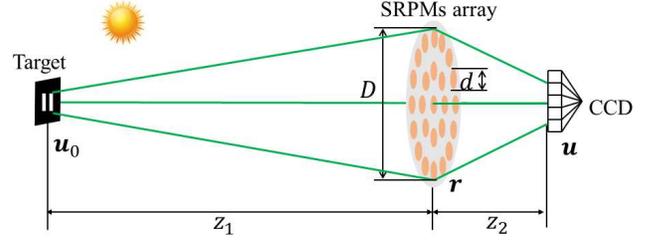}}
\caption{Schematic of wide-spectrum optical synthetic aperture imaging via spatial intensity interferometry.}
 \label{Fig1}
\end{figure}

A schematic of optical synthetic aperture imaging via spatial intensity interferometry is presented in FIG. \ref {Fig1}. An object is placed at distance $z_1$ before the sub-aperture SRPMs array. The object is illuminated by  spatially incoherent, narrowband thermal light, and a planar array light intensity detector (such as CCD and CMOS) that is placed at a distance $z_2$ behind the sub-aperture SRPMs array records the spatial intensity distribution of light that diffuses through the sub-aperture SRPMs array. Specially, the pupil function of synthetic aperture of the sub-aperture SRPMs array is expressed as
\begin{equation}
\label{eq:P}
P(\boldsymbol{r}) = a(\boldsymbol{r})s(\boldsymbol{r})
\end{equation}
where $s(\boldsymbol{r})$ is the modulation phase transfer function of the sub-aperture SRPMs array,
\begin{equation}
\label{eq:a}
 a(\boldsymbol{r})=\sum_{n=1}^N {a_{\text{sub}}^{(n)}(\boldsymbol{r})} \otimes {\delta(\boldsymbol{u}-\boldsymbol{r}_n)}
\end{equation}
is the aperture function, $a_{\text{sub}}^{(n)}(\boldsymbol{r})$ and $\boldsymbol{r}_n$ are the aperture function and the position coordinates of sub-aperture SRPM $n$, respectively. $N$ is the number of sub-aperture SRPMs, and $\delta (\cdot)$ is the Dirac function.

In the field of view (FOV) within the memory-effect range (i.e., space translation invariance) \cite{Feng1988, Osnabrugge2017, Wang2020b}, the detected intensity distribution is
\begin{equation}
\label{eq:It}
  I_t(\boldsymbol{u})=\int_{-\infty}^{\infty} I_0(\boldsymbol{u}_0)h_I(\boldsymbol{u}-\boldsymbol{u}_0)d\boldsymbol{u}_0=I_0(\boldsymbol{u})\otimes h_I(\boldsymbol{u})
\end{equation}
where $I_0$ is the image of the object, $h_I$ is the incoherent intensity impulse response function, and $ \otimes $ is the convolution operation. Therefore, the spatial intensity autocorrelation of the detected light is
\begin{equation}
\begin{split}
\label{eq:GIt1}
G_{I_t}^{(2)}(\Delta\boldsymbol{u}) &\stackrel{\text{def}} = \left\langle I_t(\boldsymbol{u})I_t(\boldsymbol{u}+\Delta\boldsymbol{u})\right\rangle_{\boldsymbol{u}} =\int_{-\infty}^{\infty} I_t(\boldsymbol{u})I_t(\boldsymbol{u}+\Delta\boldsymbol{u})d\boldsymbol{u} \\
&= G_{I_0}^{(2)}(\Delta\boldsymbol{u}) \otimes G_{h_I}^{(2)}(\Delta\boldsymbol{u})
\end{split}
\end{equation}
where ${\left\langle \cdot \right\rangle}_{\boldsymbol{u}}$ is the spatial ensemble average over the coordinate $\boldsymbol{u}$,
\begin{equation}
\begin{split}
\label{eq:GI0}
G_{I_0}^{(2)}(\Delta\boldsymbol{u}) \stackrel{\text{def}}= \int_{-\infty}^{\infty} I_0(\boldsymbol{u})I_0(\boldsymbol{u} + \Delta\boldsymbol{u})d\boldsymbol{u}
\end{split}
\end{equation}
is the spatial intensity autocorrelation of object, and
\begin{equation}
\begin{split}
\label{eq:GhI}
G_{h_I}^{(2)}(\Delta\boldsymbol{u}) \stackrel{\text{def}}= \int_{-\infty}^{\infty} h_I(\boldsymbol{u})h_I(\boldsymbol{u} + \Delta\boldsymbol{u})d\boldsymbol{u}
\end{split}
\end{equation}
is the spatial intensity autocorrelation of the incoherent intensity impulse response function. After derivation(see Supplementary Material I for details), we obtain
\begin{equation}
\label{eq:GhI1}
G_{h_I}(\Delta\boldsymbol{u})
= 1 + \left| \left\{\mathfrak{F}_a \otimes \tilde{R}_s \right\}_{-\dfrac{\Delta \boldsymbol{u}}{2\lambda z_2}} \right|^2
\end{equation}
where $\mathfrak{F}_a$ represents the Fourier transform of the aperture function $a$, and (see Supplementary Material II)
\begin{equation}
\label{eq:Rs5}
\tilde{R}_s(\Delta\boldsymbol{u})
= \exp\left\{ -\left[\dfrac{2\pi(n-1)\sigma_{\eta}}{\lambda}\right]^2\left[1 - \exp \left(-\dfrac{c_2^4 \Delta\boldsymbol{u}^2}{r_c^2}\right)\right]\right\}
\end{equation}
is the reduced autocorrelation function of the modulation phase function, where $c_2 = {z_1}/{\left(z_1+z_2\right)}$.

In the optical synthetic aperture imaging via spatial intensity interferometry, we design (see Supplementary Material III)
\begin{equation}
\label{condition}
\dfrac{4(n-1)\sigma_{\eta} z_2}{r_c} \gg D
\end{equation}
then, the result of Eq.(\ref{eq:GhI1}) is
\begin{equation}
\label{eq:GhI2}
G_{h_I}(\Delta\boldsymbol{u})
\approx 1 + \left| \left\{\mathfrak{F}_a\right\}_{-\dfrac{\Delta \boldsymbol{u}}{2\lambda z_2}} \right|^2.
\end{equation}

Taking Eq.(\ref{eq:GhI2}) into Eq.(\ref{eq:GIt1}) yields
\begin{equation}
\begin{split}
\label{eq:GIt2}
G_{I_t}^{(2)}(\Delta\boldsymbol{u})
= G_{I_0}^{(2)}(\Delta\boldsymbol{u}) \otimes \left\{ 1 + \left| \left\{\mathfrak{F}_a\right\}_{-\dfrac{\Delta \boldsymbol{u}}{2\lambda z_2}} \right|^2 \right\}
\end{split}
\end{equation}

According to the Wiener–Khinchin theorem for deterministic signals (also known as the autocorrelation theorem), we obtain
\begin{equation}
\begin{split}
\label{eq:FGI}
\mathfrak{F}{\left\{G_I^{(2)}\left( \Delta\boldsymbol{u}\right)\right\}}_{\boldsymbol{r}}
= \left| \mathfrak{F} \left\{I{(\Delta\boldsymbol{u})}\right\}_{\boldsymbol{r}}\right|^2
\end{split}
\end{equation}
Now applying the Fourier transform to Eq.(\ref{eq:GIt2}) yields
\begin{equation}
\label{eq:FGIt}
\mathfrak{F}{\left\{G_{I_t}^{(2)}\left( \Delta\boldsymbol{u}\right)\right\}}_{\boldsymbol{r}}
 = \left|\mathfrak{F} \left\{I_0{(\Delta\boldsymbol{u})}\right\}_{\boldsymbol{r}}\right|^2
\mathcal{H} \left( \boldsymbol{r} \right) + C,
\end{equation}
where $C =\left. \left\{ \left| \mathfrak{F} \left\{I_0(\Delta\boldsymbol{u})\right\}_{\boldsymbol{r}}\right|^2 \right\} \right|_{\boldsymbol{r}=0}$ is a constant, and
\begin{equation}
\begin{split}
\label{eq:H1}
\mathcal{H} \left( \boldsymbol{r} \right)
&= \mathfrak{F} \left\{ \left| \left\{\mathfrak{F}_a\right\}_{-\dfrac{\Delta \boldsymbol{u}}{2\lambda z_2}} \right|^2 \right\}_{\boldsymbol{r}}\\
&= \int_{-\infty}^{\infty} a(\boldsymbol{r_0})a\left(\boldsymbol{r}_0 + \dfrac{\boldsymbol{r}}{2\lambda z_2}\right) d\boldsymbol{r_0}
\end{split}
\end{equation}
is the autocorrelation of the synthetic aperture $a$, and the fact of $a$ is an even function has been used to replace $a\left(-\boldsymbol{r}_0 - \dfrac{\boldsymbol{r}}{2\lambda z_2}\right)$ by $a\left(\boldsymbol{r}_0 + \dfrac{\boldsymbol{r}}{2\lambda z_2}\right)$.

Thus, the diffraction-limited effects of the optical synthetic aperture imaging expressed in the frequency domain. Eq.(\ref{eq:FGIt}) shows the frequency spectra’s modulus $\left| \mathfrak{F} \left\{I_0{(r)}\right\}\right|$ of object’s diffraction-limited image intensity $I_0$ can be estimated by the intensity autocorrelation $G_{I_t}^{(2)}\left( r\right)$ of detected light intensity $I_t$, and then $I_0$ is reconstructed by phase retrieval algorithms. As shown in Eq.(\ref{eq:H1}), the frequency response of the optical synthetic aperture imaging is similar to that of the incoherent imaging case, and $\mathcal{H}$ is similar to the optical transfer function (OTF) of the incoherent imaging system. Therefore, the resolution of the object' is primarily determined by the baseline length $D$ of the synthetic aperture. And according to the Rayleigh criterion of resolution, the angular resolution is limited to $1.22 \lambda/D$ with a circular synthetic aperture \cite{Saha2010}.

To simplify the theoretical analysis, the aperture function of each sub-aperture SPRM is a circular aperture with the same diameter $d$, i.e., $a_{\text{sub}}^{(n)} (\boldsymbol{r})=a_{\text{sub}} (\boldsymbol{r})$, and
\begin{equation}
\begin{split}
\label{eq:a2}
a(\boldsymbol{r})
= \sum_{n=1}^N {a_{\text{sub}}(\boldsymbol{r})}
\otimes {\delta(\boldsymbol{r} - \boldsymbol{r}_n)}
\end{split}
\end{equation}
Substituting this into Eq.(\ref{eq:H1}), yields,
\begin{equation}
\begin{split}
\label{eq:H2}
\mathcal{H} \left( \boldsymbol{r} \right)
= \dfrac{1}{N}\sum_{n=1}^N \sum_{m=1}^N
\mathcal{H}_{\text{sub}}\left( \boldsymbol{r}-\dfrac{\Delta \boldsymbol{r}_{mn}}{\lambda z_2}\right),
\end{split}
\end{equation}
where $\Delta \boldsymbol{r}_{mn} = \boldsymbol{r}_m - \boldsymbol{r}_n$, and
\begin{equation}
\begin{split}
\label{eq:Hsub}
\mathcal{H}_{\text{sub}} \left( \boldsymbol{r} \right)
= \int_{-\infty}^{\infty} a_{\text{sub}}\left(\boldsymbol{r}_0\right)
a_{\text{sub}}\left(\boldsymbol{r}_0 +\dfrac{\boldsymbol{r}}{\lambda z_2} \right) d\boldsymbol{r}_0
\end{split}
\end{equation}
is the sub-aperture’s OTF. Eq.(\ref{eq:H2}) represents the relationship between $\mathcal{H}$ and  $\mathcal{H}_{\text{sub}}$, which is consistent with conventional synthetic aperture imaging. Finally, note that the phase error $\phi_n$ of the sub-aperture can be placed in the modulation phase transfer function $s(\boldsymbol{r})$, and its effect is primarily reflected in its influence on the statistical characteristics of $s(\boldsymbol{r})$.

Compared with stellar intensity interferometry using temporal intensity correlations, which severely limits the imaging bandwidth and sensitivity, a major advantage of the presented approach is that it does not require measuring temporal correlations of light intensity with different arrival times between photons recorded in different telescopes, and the imaging bandwidth is limited by chromatic dispersion, not by recording the temporal intensity fluctuations of light. The sensitivity of wide-spectrum optical synthetic aperture imaging is significantly improved with this approach.

\begin{figure}[htbp]
\centerline{\includegraphics[width=1.0\columnwidth]{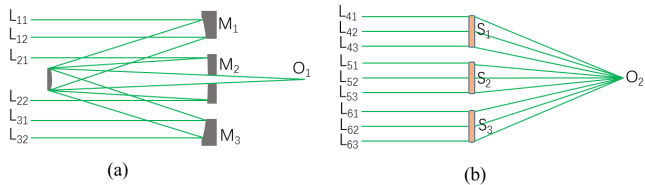}}
\caption{ (a) Traditional optical synthetic aperture system.
          (b) Schematic of wide-spectrum optical synthetic aperture imaging via spatial intensity interferometry.}
 \label{Fig2}
\end{figure}

In traditional optical synthetic aperture systems, if we seek to achieve the imaging resolution by synthetic aperture \cite{Saha2010}, not only the optical path difference $\Delta l$ between different lights L$_{11}$ and L$_{12}$ of a single aperture must be less than $1/4\lambda$, but also the optical path difference between sub-apertures M$_1$, M$_2$, M$_3$ must be less than $1/4\lambda$, as shown in FIG \ref{Fig2}. To achieve the imaging resolution limit of a very large telescope interferometer (VLTI) system \cite{Derie2000},the optical path difference between sub-apertures are controlled below $1/ 20\lambda$ using delay lines. But in the method presented in this study, the optical path difference between sub-apertures S$_1$, S$_2$, and S$_3$ can also be included in $s(\boldsymbol{r})$. The theoretical analysis shows that the imaging resolution is constrained by the baseline length of the synthetic aperture.

The imaging structure of this approach is reminiscent of a synthetic marginal aperture with revolving telescopes (SMART) \cite{Bulbul2018}, however, unlike SMART, it does not require a long invasive calibration procedure involving a guide-star to measure the system impulse response and is insensitive to atmospheric turbulence. Similar to stellar speckle interferometry, single-shot imaging through scattering media and the Wiener-Khinchin telescope, the object's image is recovered from spatial intensity autocorrelation. However, in the present approach the autocorrelation of an object's image is obtained from a single image via spatial ensemble averaging, wherea stellar speckle interferometry obtains the autocorrelation of an object's image from multiple images via temporal ensemble averaging. Compared with single-shot imaging through scattering media and the Wiener-Khinchin telescope through a monolithic spatial random phase modulator, this technique is modulated by the sub-aperture SRPMs array and its diffraction is limited by the length of the baseline of the sub-aperture SRPMs array. The angular FOV of this approach is limited by the memory effect range through the sub-aperture SRPMs array, which is inversely proportional to the SPRMs' height standard deviation of the sub-aperture array \cite{Liu2019}.

\section*{Simulation and Experiment}
\begin{figure}[htbp]
\centerline{\includegraphics[width=1.0\columnwidth]{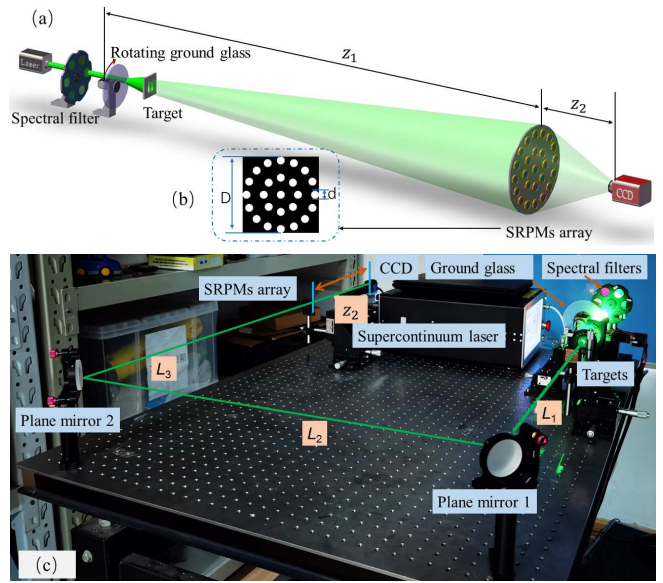}}
\caption{ (a) Schematic diagram of the simulation structure.
          (b) Structure of sub-aperture SRPMs array.
          (c) Optical path structure of the experiment.}
 \label{Fig3}
\end{figure}

A schematic diagram of the simulation is shown in FIG. \ref{Fig3}(a). The distance $z_1$ from the target to the SRPMs array was 2.00 $m$. The distance $z_2$ from the SRPMs to CCD was 0.339 $m$. The structure of the SRPMs array is shown in FIG. \ref{Fig3}(b). The diameter $d$ of the sub-aperture was 1 $mm$. The length of the baseline of the synthetic aperture $D$ was 10 $mm$. Independent spatial random phase modulators were used for each sub-aperture. In the simulation, the number of CCD pixels was $512 * 512$, and the pixel size was 4.4 $\mu m$.

The optical path structure of the experiment is shown in FIG. \ref{Fig3}(c). The main parameters of the experiment were consistent with those of the simulation, except that two mirrors were inserted into the optical path, where $L_1 + L_2 +L_3= z_1$. To produce a light source with good directionality, light was generated using a supercontinuous laser through rotating glass in the experiment. The pseudothermal light generated by this method has the same characteristics as real thermal light \cite{Liu2019}. The spectral width of the supercontinuum laser ranged from 430 to 2400 $nm$. The size of the laser spot on the ground glass surface was approximately 4 $mm$. The speed of the rotating ground glass was approximately 120-180 $r/min$. The number of pixels for the sampling CCD were 1600 $*$ 1200, $512 * 512$ pixels in the central area are selected for reconstruction calculation.

\begin{figure}[htbp]
\centerline{\includegraphics[width=0.9\columnwidth]{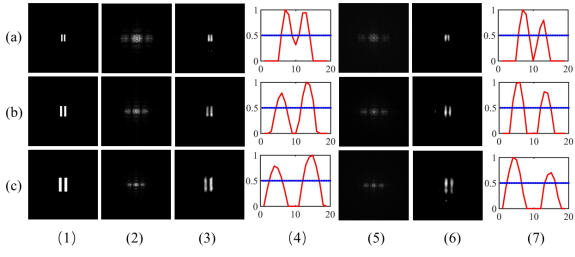}}
\caption{ Simulation and experimental results.
The spectral widths of the filters were $532\pm 0.5$ $nm$. (a1)-(c1) Targets. (2)-(4) Simulation results. (5)-(7) Experimental results. In the experiment, the exposure times of CCD were 1.2 $s$, 0.75 $s$ 0.5 $s$, and the gains of CCD were 30dB, 30dB, 28dB, respectively. (2), (5)Spatial intensity autocorrelation of CCD. (3), (6) Reconstruction of target image using phase retrieval algorithms HIO. (4), (7) One-dimensional normalized date of double slits reconstruction image. The blue lines indicate half of the maximum value.}
\label{Fig4}
\end{figure}

To test the imaging resolution, we selected double slits as the target in the simulation and experiment.The center distances $\Delta x$ of the double-slit were 130, 195 and 260 $\mu m$, respectively; they were 1.0, 1.5 and 2.0 times the diffraction limit ($1.22 \lambda z_1/D $ ) of the synthetic aperture, respectively (FIG. \ref{Fig4}(a1)-(c1)). FIG. \ref{Fig4}(a4) and (a7)show the one-dimensional normalized date of double slits reconstruction images, the results show that the resolution limit of simulation and experiment both consistent with the theory. When the distances of the double-slits were wider than the diffraction limit, they could be clearly distinguished, as shown in FIG. \ref{Fig4}(b4), (c4), (b7), (c7). Therefore, the optical synthetic aperture using a sub-aperture SRPMs array has the same imaging resolution as the full aperture system.

\begin{figure}[htbp]
\centerline{\includegraphics[width=0.9\columnwidth]{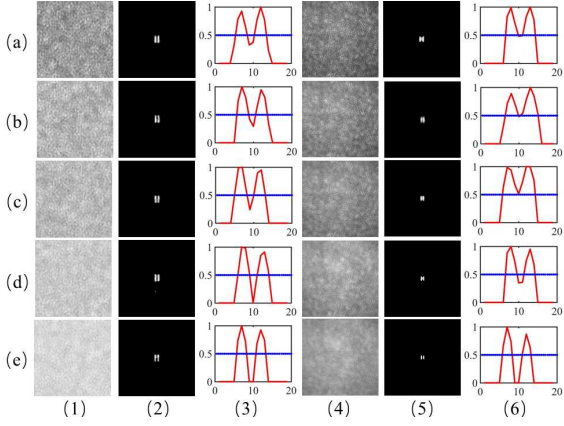}}
\caption{ Simulation and experimental results of wide-spectrum optical synthetic aperture imaging via spatial intensity interferometry.
(a)-(e) The spectral ranges of the filters were 532 $\pm 5$ $nm$, $\pm 10$ $nm$, $\pm 15$ $nm$, $\pm 25$ $nm$, $\pm 50$ $nm$, respectively.
(1)-(3) Simulational results. (4)-(6) Experimental results. The sampling exposure times of CCD were 250 $ms$, and the gains of CCD were 9, 16, 23, 25, 30 $dB$, respectively.
(1), (4) Single-shot imaging detected by CCD.
(2), (5) Reconstruction of target image using phase retrieval algorithms.
(3), (7) One-dimensional normalized date of double slits reconstruction image. The blue lines indicate half of the maximum value.}
 \label{Fig5}
\end{figure}

Optical synthetic aperture imaging via spatial intensity interferometry has another advantage: it can realize wideband spectrum optical intensity interference imaging. The simulation and experimental results are shown in FIG. \ref{Fig5}. We used spectral filters to select the wavelengths of the light source. The spectral ranges of the filters were 532 $\pm 5$ $nm$, $\pm10$ $nm$, $\pm 15$ $nm$, $\pm 25$  $nm$, $\pm 50$ $nm$, respectively. The target was a double-slit whose center distance $\Delta x$ of slits is 130 $\mu m$. The single-shot imaging detected using the CCD is shown in FIG. \ref{Fig5}(1), (4); the contrast of images decreased with the increase in spectral width. However, the resolution of the image increased with the spectral width, as shown in FIG. \ref{Fig5}(3), (6). This may have been due to the shorter-wavelength components with a wider spectral width. For example, when the spectral width was 532 $\pm$ 50 $nm$, the shortest wavelength was 482 $nm$. When $z_1 =$ 2.00 $m$, the imaging resolution of the system was 118 $\mu m$, which was shorter than 130 $\mu m$. Both the simulational and experimental results verified that optical synthetic aperture imaging via spatial intensity interferometry can be used for imaging with a wide spectral width. Compared with traditional coherent synthetic aperture optical intensity interference imaging, the detection sensitivity of this system is significantly improved owing to the increase in spectral width.

\begin{figure}[htbp]
\centerline{\includegraphics[width=0.9\columnwidth]{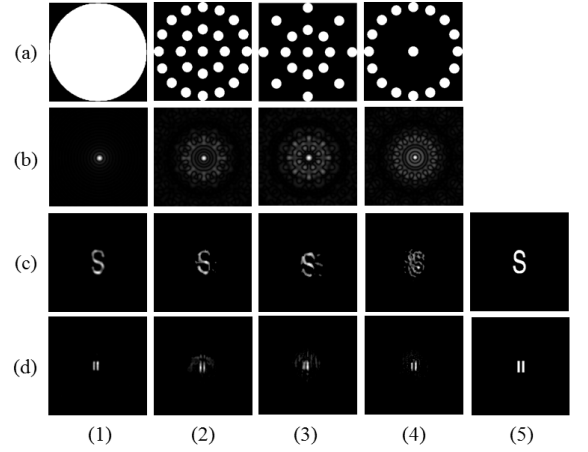}}
\caption{ Experimental results of different sub-aperture SRPMS arrays.
(a) Different sparse array structures.
(b) Fourier transform of different sparse array structures.
(5) Target letter $'s'$ and double-slit, whose sizes were 2.0 $mm$ and 0.975 $mm$, respectively.}
 \label{Fig6}
\end{figure}

A special sub-aperture structure was designed for simulation and experiment, as shown in FIG. \ref{Fig3}(b). To test the imaging effect of different sparse array structures, we designed a full aperture and other structures designed for comparative experiments, as shown in FIG. \ref{Fig6}(a). The experimental results show that the imaging resolution of the synthetic aperture system was maintained, but the imaging quality of sparse array structures was not as high as that of a full aperture structure, because synthetic apertures are sparse array structures, which have the problem of frequency spectrum loss. The Fourier transform of the full circular aperture structure is the  1th Bessel function \cite{Saha2010}, but the Fourier transform of sparse array structures are accompanied by strong sidelobes \cite{Shannon1983}, as shown in FIG. \ref{Fig6}(b). Therefore, using a sparse array structure synthetic aperture to sample has the problems of missing information and sidelobe disturbance. Therefore, the sparse array structure should be optimized according to specific objectives and application scenarios.

\section*{Discussion and Conclusion}
Optical synthetic aperture imaging via spatial intensity interferometry can overcome the high time coherence and narrow-band spectral width required by the traditional intensity interference imaging. When the diameter $D$ of the synthetic aperture is 10 $mm$ and $z_2$ is 0.339 $m$, the optical path difference between the center light L$_{52}$ and the edge light L$_{41}$ is 36.9 $\mu m$, which is equivalent to $69.36\lambda$ when the central wavelength is 532 $nm$ and much longer than $1/4\lambda$, as shown in FIG. \ref{Fig2}. Therefore, the optical synthetic aperture imaging mentioned in this study is insensitive to the optical path difference between the sub-apertures, which significantly reduces the difficulty of synthetic aperture construction. Recently, astronomers have shown the first picture of a supermassive black hole in the center of the Milky Way galaxy, which is an important proof of the resolution advantage of the synthetic aperture telescope in astronomical observation \cite{Akiyama2022}.

In addition, in the experiment and simulation, the spectral width of the system reached 100 $nm$, which significantly improved the detection sensitivity of the system. In principle, this method is not limited to the width of the spectrum, but a spectrum that is too wide will significantly reduce the contrast of the detected image using CCD. If we want to realize the effective detection of wide-band intensity interferometry, the dynamic range of the detector CCD must be sufficiently high to satisfy the requirements of data acquisition.

This study provided a possible solution to reducing the construction difficulty of optical synthetic aperture telescopes and increasing the baseline length of optical telescope to 100 $m$ or even kilometers. Wide-spectrum optical synthetic aperture imaging via spatial intensity interferometry has potential application value in astronomical observation and space target high-resolution imaging.

\section*{Supplementary Material}
See the supplementary material for the derivation of the incoherent intensity impulse response function's spatial intensity autocorrelation, the correlation of modulation phase transfer function, and the design condition of sub-aperture spatial random phase modulator.
\section*{Funding}
This research is supported by the National Natural Science Foundation of China (NSFC) NO.61991454.

\section*{Acknowledgment}
Thanks are due to Dr. Feng Lu (National Astronomical Observatory, Chinese Academy of Sciences) for valuable discussion about the optical synthetic aperture systems. 

\section*{Author Disclosures}
The authors declare no conflicts of interest.

\section*{References}
%


\end{document}